\documentclass[useAMS,usenatbib]{mn2e}

\usepackage{graphics}
\usepackage{epsfig}

\title[A Blue Tilt in NGC 5170]{A Blue Tilt in the Globular Cluster 
System of the Milky Way-like Galaxy NGC 5170}
\author[D. A. Forbes et al.]{Duncan A. Forbes$^{1}$\thanks{E-mail:
dforbes@swin.edu.au}, Lee R. Spitler$^{1}$, W. E. Harris$^{2}$, Jeremy Bailin$^{2}$, Jay Strader$^{3}$$\footnotemark[1]\thanks{Hubble Fellow}$,
\newauthor 
Jean P. Brodie$^{4}$, S. S. Larsen$^{5}$ 
\\
$^{1}$Centre for Astrophysics \& Supercomputing, Swinburne University, Hawthorn VIC 3122, Australia\\
$^{2}$Department of Physics and Astronomy, McMaster University, Hamilton ON, L8S 4M1, Canada\\
$^{3}$Harvard-Smithsonian Cetre for Astrophysics, Cambridge, MA 02138, USA\\
$^{4}$UCO/Lick Observatory, Santa Cruz, CA 95064, USA\\
$^{5}$Astronomical Institute, Utrecht University, Utrecht, 
N-3584, The Netherlands}
\begin{document}

%\date{Accepted 1988 December 15. Received 1988 December 14; in original form 1988 October 11}

\pagerange{\pageref{firstpage}--\pageref{lastpage}} \pubyear{2002}

\maketitle

\label{firstpage}

\begin{abstract}
Here we present HST/ACS imaging, in the B and I bands, of the edge-on
Sb/Sc galaxy NGC 5170. Excluding the central disk region region, we
detect a 142 objects with colours and sizes typical of globular
clusters (GCs).  
Our main result is the discovery of a `blue tilt' (a mass-metallicity
relation), at the 3$\sigma$ level, in the metal-poor GC subpopulation of this Milky Way like
galaxy. The tilt is consistent with that seen in massive elliptical
galaxies and with the self enrichment model of Bailin \& Harris. For a linear mass-metallicity 
relation, the tilt has the form Z $\sim$ L$^{0.42 \pm 0.13}$.  
We derive a total GC system population of 600 $\pm$
100, making it much richer than the Milky Way. However when this number
is normalised by the host galaxy luminosity or stellar mass it is 
similar to that of M31. Finally, we report the presence of a potential Ultra Compact
Dwarf of size $\sim$ 6 pc and luminosity M$_I$ $\sim$ --12.5, assuming
it is physically associated with NGC 5170.
\end{abstract}

\begin{keywords}
globular clusters: general -- 
galaxies:star clusters -- galaxies: individual (NGC 5170)
\end{keywords}

\section{Introduction}

Studies of the globular cluster (GC) system of the Milky Way have benefited 
from the wealth of information that is available. However such studies are 
also fundamentally limited by the small sample size of about 150 objects. 
Discerning more subtle trends in GC properties, or finding GCs in more extreme 
astrophysical states, requires larger samples which 
can only be found in the GC systems of other galaxies. 

One such trend 
which was revealed first in extragalactic GC systems, 
and not seen in the Milky Way GC system, is the so-called `blue tilt'.
The blue tilt (a trend for the blue GC subpopulation to
have redder colours at brighter magnitudes) has been found using the 
Advanced Camera for 
Surveys (ACS) on the {\it Hubble Space Telescope} (HST) in a
variety of galaxies from the most massive ellipticals (Harris et
al. 2006), to lower mass ellipticals (Strader et al. 2006) and even
dwarfs (Mieske et al. 2006a). It has also been detected in an
early-type spiral galaxy, the Sombrero, by Spitler et al. (2006).  It
has not been detected in any HST study of late-type spirals, but the
few studies to date (e.g. Goudfrooij et al. 2003) 
have generally used the WFPC2 camera which has a more limited
field-of-view compared to the ACS camera and hence such studies have 
focused on the bulge region and their associated red GCs. 
Interestingly, at this stage, it has not been detected
in the rich GC system of the massive elliptical M49 (Strader et
al. 2006; Mieske et al. 2006a).

Given the evidence that extragalactic GCs are mostly old (Brodie \& Strader 
2006), 
the blue tilt implies a mass-metallicity relation for the metal-poor 
subpopulation of GCs (although this has yet to be confirmed 
spectroscopically). Various explanations have been proposed 
(see Bekki et al. 2007; Mieske et al. 2006a; Mieske 2009), but 
perhaps the most plausible explanation to date is one of 
self enrichment, whereby more massive GCs are enriched with more  
heavier metals during their brief formation period. 

In the self enrichment models of Strader \& Smith (2008) and Bailin \&
Harris (2009), a single generation of star formation and resulting
supernovae is responsible for the self enrichment of each GC, which
only becomes important above a threshold mass of $\sim$ 10$^6$
M$_\odot$. 
%Differences lie in the assumptions regarding star formation
%efficiency and retention of metals, and how these parameters vary with
%cluster properties. As well as the threshold mass, the Bailin \&
%Harris (2009) model explicitly predicts the metallicity scatter at a
%given cluster mass.  
Both models predict heavy element abundance variations
within each massive GC, which may be the situation in some extragalactic GC
systems (e.g. Cenarro et al. 2007). The Galactic GC system includes
only a few GCs with masses $\ge$ 10$^6$ M$_\odot$, so a clear
signature of variation in colour (or metallicity) with
magnitude (a blue tilt) is difficult to detect.
It is thus important to determine whether 
galaxies similar to the Milky Way reveal a GC blue tilt or not. 

Recently the ACS was used to study the star cluster system of 
the Sc galaxy NGC 3370 
by Cantiello et al. (2009). NGC 3370 is a near face-on spiral
at a somewhat larger distance of 28.7 Mpc. These two properties make
it more difficult to study its GC system.  Cantiello et al. detected
35 GC candidates with a mean colour B--I = 1.60 $\pm$ 0.27 but no evidence of a blue tilt.

Here we present an analysis of the globular cluster system of NGC
5170 using the ACS camera on board the
HST. NGC 5170 is a potential analogue of the Milky Way Galaxy with a
Hubble type of Sb (Sandage \& Bedke 1994) to Sc (NASA Extragalactic Database; NED). Fischer et
al. (1990) derive a bulge-to-disk ratio of 0.5 which is more
consistent with an Sb, like M31, than an Sc.  It is a relatively
isolated, edge-on ($i$ $\sim$ 86$^o$; Bottema et al. 1987) 
spiral with a total V band magnitude of 9.84
(NED). It lies at a Galactic latitude of 43$^{o}$, making the surrounding 
field relatively free of Galactic stars and hence a good target for 
a GC study. 
A wide range of distance estimates are found in the
literature. Ferrarese et al. (2000) report the GC luminosity function
distance of Fischer et al. (1990) of 21.5 Mpc.  However, Fischer et
al. did not reach beyond the GC turnover magnitude so this value
should be considered somewhat uncertain.  The Tully-Fisher based
distance estimates of Willick et al. (1997) range from 24 to 43 Mpc.
The kinematic study of NGC 5170 by Bottema et al. (1987) assumed a distance of 
20 Mpc. 
The Virgo infall corrected velocity (NGC 5170 lies behind but close
to the Virgo cluster southern extension) from NED is 1665 $\pm$ 27 
km $^{-1}$, which with 
%(as done by Kregel \& van der Kruit 2005), 
H$_o$ = 73 $\pm$ 5 km s$^{-1}$ Mpc$^{-1}$ corresponds 
to a distance of 22.8 $\pm$ 1.6 Mpc (m--M = 31.79 $\pm$ 0.15). The final error 
is the quadrature sum of the velocity error combined with the error on the Hubble constant. 
For this distance, the galaxy has an absolute magnitude of 
M$_V$ = --21.95.  
We initially adopt this latter value, for which the 
pixel scale of the ACS (0.05$^{''}$) equals $\sim$5 pc, but we reinvestigate
this issue below and derive a new distance estimate based on measured sizes of the NGC 
5170 GCs. 

\section{Observations and Data Reduction}

The data for this work was taken using the ACS/WFC camera on board the
HST as part of proposal ID 9766 (PI
Forbes). Two pointings along the major axis of NGC 5170 were obtained
in the F435W (B) and F814W (I) filters with a total exposure time of
3720s and 940s at each pointing. A $\sim$500 pixel overlap between
the two pointings gives a final area coverage of about 370$^{''}$ $\times$ 210$^{''}$.
This corresponds to roughly 40 kpc $\times$ 23 kpc (for an assumed distance of 22.8 Mpc), which 
would emcompass over 90\% of the Milky Way's GC system. 
The inter-chip gap region is projected along the
disk of the galaxy.

Initial data reduction was carried out using the standard ACS
calibration pipeline.  Multiple exposures were aligned and combined
using the multidrizzle task within IRAF.  The profiles of common stars
in the overlap region were used to check the alignment
process. The final two-pointing image is shown in Fig. 1.

\section{Globular Cluster Selection and Photometry}

Before the initial object detection, we subtracted a smooth model of
the galaxy light from the image, using the ELLIPSE task in IRAF, to help with object detection and
photometric analysis. DAOFIND was then used to select objects with a
threshold 4 times the standard deviation of the background count level
for both the B and I images.  

An aperture magnitude was measured on
all objects using an extraction radius of 5 pixels and a local
background level estimated with a 5 pixel wide annulus starting at 20
pixels from the object.
%Size-dependent aperture corrections derived in Harris~(2009) were used
%to account for excess light falling between 5 and 20 pixels.  

The
angular sizes of objects were determined with the latest version of
the ISHAPE (Larsen 1999) program.  We used 
empirical PSFs, which were constructed for the B and I band mosaics
using $\sim30$ point sources across the images.  Point sources were
identified as such using the FWHM calculated from a pass of SExtractor
(Bertin \& Arnouts 1996) over each mosaic.  Isolated star-like objects with FWHM of
1.8--2.0 pixels (see Harris 2009a) were used to construct the empirical
PSFs for ISHAPE with the standard DAOPHOT {\it PSF} software routine.
Again following Harris (2009a), we used a King30 light profile
with the default concentration parameter of $c=1.5$ (King 1962).
The ellipticity and position angle
were allowed to vary during this fit.  The final half light radii
(r$_{h}$) were used to pick an appropriate aperture correction to derive total magnitudes 
following Harris (2009a). 

Before converting the photometry to a standard photometric system,
Galactic extinction corrections were applied using E(B--V) $=0.08$
from the DIRBE dust maps (Schlegel et al.~1998) and extinction
coefficients for a G1-star from Sirianni et al. (2005). 
No internal
reddening correction is applied.  Conversion of the total instrumental
magnitudes to the standard Johnson B and I bands were carried out
using the procedure of Sirianni et al. (2005) including colour terms.
Charge transfer efficiency corrections are likely to be negligible in
the bright galaxy halo regions of our detected objects, and were not
applied. The B and I band objects were matched in position (with a 3 pixel tolerance) 
to create a common object list of sizes 
and extinction-corrected photometry.

Objects projected on the disk and inner bulge are very difficult to 
disentangle from star clusters and OB associations, and are problematic when 
it comes to measuring reliable photometry and sizes. We opted to 
apply a spatial 
cut to exclude the disk and inner bulge regions. This cut essentially 
removes a region above and below the (edge-on) disk.
%, which grows in 
%vertical size. 
It means we may have partially selected against red (metal-rich) 
bulge/disk GCs that may be present, which preferentially lie in these regions. 

%Resolved background galaxy contamination was identified and
%subsequently removed when an object's measured ISHAPE ellipticity
%($\epsilon=b/a$) in either band was less than
%$\epsilon+\Delta\epsilon<0.7$, where $\Delta\epsilon$ is the
%cross-correlated errors on $\epsilon$ computed by ISHAPE.  Each band
%was considered separately since an elongated structure is sometimes only
%apparent in one band.

As the distance to NGC 5170 is quite uncertain, we next applied a 
(distance-independent) colour selection.  
We selected only those objects with colours $1.2 <$ (B--I)$_o$ $\le 2.2$.
Such a selection applied to the Milky Way would include over 95\% 
of the GC system (Harris 1996). 

Harris (2009a) has shown that objects with an ISHAPE signal-to-noise ratio of 
less than 50, have unreliable size measures. Our data support this conclusion. 
We have therefore decided to select only those objects 
with an I band signal-to-noise 
ratio greater than 50. As shown in Fig. 2, 
the signal-to-noise ratio tracks the magnitude 
of an object, so that this selection cut is effectively a magnitude 
limit of I $\le$ 23 (which is a few tens of a magnitude brighter than the expected 
GC turnover magnitude, as discussed below). 
We note that 
the B band signal-to-noise ratio closely follows that of the I band 
(as the exposure 
times were chosen to give similar effective depths).

%removed 6 objects brighter than I = 19.1, 5 stars and 1 large elliptical gal.
%would be picked up in visual check or size check.

At this stage in our selection process, our object list contains several 
background galaxies. 
%Examination of the objects indicated that our 
%ellipticity measure was not robust enough to be used a selection criterion. 
%Below we focus on object size, in the I-band, 
The next step in our selection criteria is to use the size as measured by 
ISHAPE as a means of excluding 
extended background sources and for identifying {\it bona fide} GCs.

The combination of high spatial resolution imaging from the ACS camera 
and the high
Galactic latitude of NGC 5170 ensures that the contamination from
foreground stars in our final object list is minimal. 
Nevertheless we applied a {\it minimum} size selection
removing $\sim$50 objects with half light sizes less 
than 0.01$^{''}$ as these are most likely stars (see Harris 2009a).  
%removed 54 objects with this step.

%\subsection{Globular Cluster Size Selection}

The half light sizes of GCs are thought to remain relatively stable 
as a GC evolves (Aarseth \& Heggie 1998) and are found observationally 
to have a near constant average 
size in large galaxies (Jordan et al. 2005; Harris et al. 2009a). The latter 
suggests that comparison of the physical sizes of the objects in NGC 5170 with 
those of the Milky Way can be used to further refine our object list. 
%In a sample of 92 well-observed 
%Milky Way GCs with I band magnitudes and sizes, the 
%median half light size is 2.8 pc (Harris 1996).
%Correcting to a standard galaxy colour, projected radius and 
%GC colour, Jordan et al. (2005) have measured a mean half light size of 
%2.7 $\pm$ 0.3 pc. 
%Using equation 21 from their paper, one can apply a rough 
%correction to the measured mean sizes for NGC 5170 to obtain a new distance 
%estimate to NGC 5170. Here we have assumed 
%that the projected radii are typically distributed at twice 
%the effective radii, that the galaxy has a g--z colour typical of an 
%Sbc spiral and that our GCs are dominated by the 
%metal-poor subpopulation with a metallicity of [Fe/H] = --1.7 or (g--z) = 
%0.91. The correction is dominated by the galaxy and GC colour terms, and 
%results in a correction factor of $\sim$0.75.

%We next proceeded to apply a series of {\it maximum} size selections.
%A visual check of large sized objects revealed that a size cut of
%0.1$^{''}$ is quite conservative and unlikely to remove any 
%{\it bona fide} GCs. 
%Our sample at this stage contains 254 objects. 
%The median size after this cut is applied is 0.032$^{''}$ (3.5 pc). 
%In a Milky Way GC sample, this size selection would exclude 
%two GCs.  
 
In Fig. 3
we show the size-magnitude distribution of objects assuming a Virgo-infall 
distance 
of 22.8 Mpc. The plot shows a fairly-well defined locus of objects in NGC
5170 with median sizes $\sim$ 3.5 pc, which appear to have slightly larger sizes
(and to be more luminous) than the bulk of the Milky Way GC system. A closer distance 
to NGC 5170 would provide better agreement between the two distributions. 
The brightest GC in the Milky Way, Omega Cen, with an
extinction-corrected magnitude of M$_I$ = --11.2 (and (B--I)$_o$ =
1.56) is labelled. The second brightest is M54 with M$_I$ = --10.9
(and (B--I)$_o$ = 1.61).
In Fig. 4 we 
show the same data but adjust the distance to NGC 5170 to be 19.5 Mpc (this choice of 
distance is justified below). Visually the two distributions appear better matched. 
%Fig. 4 also shows that the brightest GCs of the NGC
%5170 GC system are similar in magnitude to that of the Milky Way.
%(We note that the brightest Milky Way GCs may be better regarded as
%the remnants of dwarf galaxy nuclei.)

In an effort to create a cleaner sample (i.e. reduce the number of
background galaxy contaminants) we imposed a size limit of r$_h$ $<$
0.05$^{''}$ (4.7 pc for a distance of 19.5 Mpc). The effect of this
selection can be seen in Fig. 4. It shows that half a dozen Milky Way
GCs (including Omega Cen) would be excluded, but the vast bulk
remain. We note that the Milky Way GC system, and other galaxies,
reveal a size versus galactocentric distance trend, such that the
larger GCs tend to be located at large distances (van den Bergh et
al. 1991; Jordan et al. 2005; Spitler et al. 2006; Gomez \& Woodley
2007; Harris 2009b).  The most distant selected object in our sample is
at a projected radius of $\sim$15 kpc, compared to the Milky Way GC
system which extends to $\sim$100 kpc. Thus a stricter size selection
for NGC 5170 is probably justified.
%After this size cut we have
%**193** objects remaining with a median size of r$_h$ = 0.028$^{``}$ (2.6
%pc).  We note that the UCD, with a size of r$_h$ = 0.065$^{''}$, 
% is no longer present in the r$_h$ $<$ 0.05$^{''}$ sample. 

Finally, we conducted a visual inspection of the selected
objects.  This was performed independently by two people (DF and
JS). We attempted to identify objects that we could confidently
exclude as being non-GCs on the basis of their visual appearance. We
identified 11 (DF) and 15 (JS) objects. There were 9 objects in common
and these were generally classified as small galaxies. The different
classifications tended to occur for the faint objects located near the
galaxy disk, i.e. they may be reddened OB assocations. We have chosen
to exclude the combined list of 17 objects.  The
small number of visually excluded objects, several of
which are at the faint end of our sample, do not affect our conclusions in
terms of the colour-magnitude diagram, luminosity function or total population
calculations. Our final object list contains 142 candidate GCs. Positions and 
photometry for these candidate GCs are listed in an on-line table. 

The distance of 19.5 Mpc used above, was based on the empirical evidence that 
the average half light radius for GCs is relatively constant from system to system 
(Jordan et al. 2005; Harris et al. 2009a). To compare the NGC 5170 and Milky Way GC 
systems, we first restricted the Milky Way system to be only those GCs which have M$_I$ $<$ --8.5 
(our effective magnitude limit) and we removed the 
large, outer GC NGC 2419 with a size of 17.9 pc. The remaining 56 GCs have a median half light 
radius of 3.12 pc with an error of $\pm$0.20. 
The median half light radius for the final 142 NGC 5170 candidate GCs is 
0.033$^{''}$ $\pm$ 0.001. This angular size 
matches the Milky Way physical size for a distance of 19.5 $\pm$ 1.4 Mpc (m--M =
31.45). This value is within $\sim$2$\sigma$ of the Hubble flow distance of 22.8 $\pm$ 1.6 Mpc. 
%The effect of $\pm$ 2.5 Mpc is shown in the plot by a diagonal line.  
A histogram of the NGC 5170 candidate GCs 
(for a distance of 19.5 Mpc) compared to the Milky Way GC system 
is shown in Fig. 5. They have the same median half light size. 

%In an attempt to test the robustness of our results to the size
%selection criterion, we also selected a subsample of objects with
%r$_h$ $<$ 0.035${''}$ (3.9 pc). This subsample includes the main locus
%of NGC 5170 points in Fig. ** and contains 142 objects.

Fig. 6 shows the spatial distribution of the candidate GCs within our 
field-of-view of $\sim$28 $\times$ 18 kpc (for a distance of 19.5 Mpc). 
They are concentrated around 
the galaxy centre indicating that they are indeed associated with NGC 5170.
%The spatial distribution of 
%objects with sizes r$_h$ $<$ 0.05$^{''}$
%(5.5 pc) and the subsample with r$_h$ $<$ 0.035$^{''}$ (3.9 pc) are
%shown in Fig. **. There is no significant differences in the main
%sample and the high quality subsample; both show a concentration
%towards the galaxy centre. 
Our spatial selection avoiding the region $\sim$ $\pm$ 1 kpc above and
below the disk of the galaxy is clearly seen. A similar region in the Milky Way 
would exclude about 1/3 of the GCs and 
the vast bulk of starforming regions as the scale height for HII regions 
is $\sim$50 pc (Paladini, Davies \& de Zotti 2004). 
%A comparison of objects
%with sizes r$_h$ $<$ 0.05$^{''}$ (4.7 pc) and the subsample with r$_h$
%$<$ 0.035$^{''}$ (3.3 pc) showed 
We found no significant trend in object size or magnitude with spatial
location. The GCs located near to the galaxy bulge tend to have red
colours, otherwise there is no strong colour dependence with spatial
location. 
%The location of the possible UCD which lies SE of the galaxy
%disk is also shown.
We have identified a possible Ultra Compact Dwarf (i.e. a
GC-like object with a large size and high luminosity) as labelled in
the figure. We discuss this UCD object in more detail in Section 6 below.

\section{Colour-magnitude diagram}

The observed colour-magnitude diagram for candidate GCs 
%objects with sizes r$_h$ $<$ 0.05$^{''}$ (5.5 pc and the subsample 
%with r$_h$ $<$ 0.035$^{''}$ (3.9 pc) 
is shown in Fig. 7. We note that 
%although we have selected on GC size, 
the location of objects in the colour-magnitude diagram is distance-independent.  The
colour-magnitude diagram reveals a concentration of blue objects with colour (B--I)$_o$
$\sim$ 1.6. Such colours are associated with the metal-poor
subpopulation of GCs (e.g. Barmby et al. 2000). 
It can been seen that the general locus of this blue subpopulation
for I $\le$ 22 tends to have redder colours at brighter
magnitudes; i.e. a blue tilt.  
%It is unclear whether the tilt continues to fainter magnitudes.
The metal-rich subpopulation is expected at (B--I)$_o$ $\sim$
1.9. There are a few objects with such colours, but as noted earlier
we partially selected against bulge/disk GCs which comprise the
metal-rich GC subpopulation in spirals. Within our defined
field-of-view, and assuming a blue/red colour division at (B--I)$_o$
$\sim$ 1.75, we find the fraction of red-to-blue GCs to be
0.29. Extending to larger galactocentric radii would tend to
decrease this fraction, while accounting for missing bulge/disk GCs
would tend to increase it.  However, assuming a spatial distribution
similar to the Milky Way's GC system, this fraction will be largely
unchanged at 1/3.

%Although we have excluded the region $\pm$1 kpc above and below the disk, 
%there is still a concentration of sources towards 
%the blue/disk region.  
%We find that the colour-magnitude diagram and blue tilt are robust to this effect when 
%we extend the region of excluded GCs to $\pm$2 kpc. 

%Nearby
%background, late-type spirals have colours in the range 1.5 $<$ B--I
%$<$ 2.0 and despite our size selection may make a small contribution
%to the objects at large galactocentric radii. However, as Fig. 8 shows the 
%candidate GCs are concentrated towards the galaxy centre indicating that 
%they are indeed associated with NGC 5170. 

%By matching the NGC 5170 object half light size distribution to that
%of the Milky Way, we can estimate a revised distance to NGC 5170
%(assuming GC average sizes are similar between the two galaxies).
%From this comparison we estimate a distance of 19.5 $\pm$ 2.5 Mpc, where the
%uncertainty represents our confidence limit.  Fig. ** shows the
%revised size vs luminosity distribution for a distance of 19.5 Mpc.
%At this distance, the median size of 0.028$^{''}$ corresponds to 2.64
%pc.

In an attempt to quantify the visual trend for the blue subpopulation to 
become redder with increasing brightness, we have used the NMIX statistical 
test (see for example Spitler et al. 2006). We divide 
the sample at I = 22, which corresponds to roughly half of the sample in 
number. For the brighter half, we find 42
(63\%) of the GCs are assigned to the blue subpopulation 
with a mean colour, rms dispersion and  error on the mean (B--I)$_o$ = 1.580 $\pm$ 0.101
(0.016) and 24 (37\%) are red with values  
of (B--I)$_o$ = 1.828 $\pm$ 0.151 (0.031). 
For the fainter half, the blue/red 
fractions are similar at 53 (70\%) to 23 (30\%).
However the mean colour, rms dispersion and  error on the mean are now (B--I)$_o$ = 
1.516 $\pm$ 0.115 (0.016) and 1.771 $\pm$ 0.172 (0.036). 
Thus there is a marginal 2$\sigma$
difference in the mean colour of the red subpopulation with magnitude but the bright 
blue GCs are redder than the faint blue ones at the $\sim$3$\sigma$ level.
In Fig. 7 we overlay the mean 
value, error on the mean and the rms dispersion for the blue subpopulation.  
The colour difference is $\Delta$(B--I)$_o$ = 0.064 $\pm$ 0.02.
If we assume a linear transformation from colour to metallicity of the form 
$\Delta$(B--I)/$\Delta$[Fe/H] = 0.375 (Harris et al. 2006), then this translates into 
Z $\sim$ L$^{0.42 \pm 0.13}$. 
A further division into 
three magnitudes bins supports the colour-magnitude trend of the blue 
subpopulation but with lower significance. Thus we confirm the visual 
trend in Fig. 7 of colour bimodality with a blue tilt and 
no strong evidence of a red tilt. 
We note that the appearance of the blue
tilt does not depend strongly on the exact choice of a maximum size criterion 
nor on the region excluded above and below the disk. 
%Objects with sizes 3.9 $<$ r$_h$ $<$ 5.5 pc tend to located at faint
%magnitudes suggesting they are indeed less likely to be true GCs.  

In Fig. 8 we show the colour-absolute magnitude diagram for a
distance of 19.5 Mpc (m--M = 31.45). The universal peak of the GC
luminosity function occurs at M$_I$ $\sim$ --8.3, which is fainter than the
cutoff magnitude of our data. Thus if
this is the true distance, we can not use our observed GC luminosity
function to accurately determine the distance.

Fig. 8 also shows a comparison 
to the well-defined blue tilt of Harris (2009a) who used a composite GC system 
from several massive early-type galaxies. The data 
were also taken in the B and I bands with the 
ACS camera. We show the 
mean values of the blue and red GC
subpopulations and his 2nd order polynomial fit to the blue subpopulation. 
Although they extend to higher luminosities, the blue tilt
observed in these massive early-type galaxies appears to be similar to
that seen in NGC 5170 for the range --9.5 $>$ M$_I$ $>$ --11, scattering 
evenly about the Harris fit line. Harris (2009a) quotes the mass-metallicity relation for a 
simple linear fit to the massive early-type galaxies of Z $\sim$ L$^{0.48 \pm 0.12}$. 
This is consistent with that found for NGC 5170 candidate GCs (over a more limited magnitude range).

There is an indication that the metal-rich GCs 
NGC 5170 are slightly bluer (more metal-poor) on average than 
those in the massive ellipticals. This is to be expected given the 
correlation of the mean colour of the red subpopulation with host galaxy 
luminosity (Forbes, Brodie \& Grillmair 1997; Brodie \& Strader 2006).

A comparison to the Milky Way GC system is given in Fig. 9. To our
cutoff magnitude of M$_I$ $\sim$ --8.5, the two distributions are
similar in form despite the fact that extinction corrections for some Milky
Way GCs can be quite uncertain. This figure also clearly shows that
there are more candidate GCs associated with NGC 5170 than the Milky
Way. This includes the red subpopulation, for which we are incomplete
at all magnitudes.

In Fig. 10 we reproduce the NGC 5170 GC colour-absolute magnitude
diagram again, but with a simulated GC system added.  Here we
have used the self enrichment model of Bailin \& Harris (2009) to
generate a distribution of model GCs as described in Harris et
al. (2009a), incorporating early self enrichment plus later dynamical
mass loss.  The blue subpopulation was assumed to have a mean
metallicity [m/H] = --1.6 and the red ones with [m/H] = --0.7.  Each
sequence was assigned an internal metallicity spread
%$\sigma$[m/H] = 
of $\pm$ 0.35 dex.  In the figure, we show the resulting mean and rms dispersion of the 
simulated metal-poor and metal-rich GCs for these input values. The blue tilt is apparent in the simulated data 
for bright magnitudes. For a linear mass-metallicity relation 
fit to all four mean values, the resulting relation has the form Z $\sim$ L$^{0.54 \pm 0.07}$. 
This is consistent with that found for the NGC 5170 candidate GCs.  
Fig. 10 also shows the mean colours of the blue subpopulation in NGC 5170. 
%(B--I)$_o$ = 1.580 $\pm$ 0.016 
%for bright NGC 5170 GCs compared to 1.516 $\pm$ 0.016 at fainter magnitudes. 
%while the large open symbols show the real
%NGC 5170 data.  
%We conclude from this test that the basic
%precepts of the Bailin/Harris 
We note that the Bailin \& Harris (2009) model 
also captures, to first order,  
the main features of the colour-magnitude diagram of 
the Sombrero galaxy GCs (Harris et al. 2009a) and GCs in 
other massive ellipticals (Wehner et al. 2008; Harris 2009b).

\section{Estimating the total GC system population}

\subsection{Spatial Distribution}

We follow the method of Kissler-Patig et al. (1999) and Harris et
al. (2009b) to estimate our spatial coverage of the NGC 5170 GC system. 
This method assumes that the Milky Way and NGC 5170
GCs have a similar spatial distribution.
%We estimate the number of Milky Way GCs that would be
%projected within our search area for a distance of 19.5 Mpc. 
Globular
clusters within 1 kpc of the Milky Way centre were excluded and the Y
(parallel) and Z (perpendicular) coordinates were then projected onto
our search area.   We found 70 out of 150 GCs 
in the catalogue of Harris (1996) are contained in our
search area. Thus spatially we detect almost half of the total GC system.
%a XZ search gave a few percent less.  
%This number is quite sensitive to the  
%disk region excluded perpendicular to the disk. For example, if  we exclude 
%a $\pm$2 kpc region about the disk axis 
%then the number of projected Milky Way GCs reduces to 33/151 (22\%).  

If NGC 5170 were located at a distance 10\% further (closer) then
the number found would decrease (increase) by about 4 (3\%) GCs. The change
in the number of GCs is relatively insensitive to distance as any
increase in the area of the halo searched is compensated for by the size
of the disk region that is excluded.

\subsection{Completeness Tests}

In order to estimate the magnitude completeness of our sample, we added 
simulated point sources to the B and I band ACS images (the actual GCs in NGC 5170 
are partially resolved with half light sizes of $\le$ 1 pixel). 
These simulated 
sources were fully calibrated onto the standard Johnson photometric system 
in a similar manner to the actual GC candidates with an intrinsic colour 
of B--I = 1.7 imposed. 
We ensured that none of the simulated sources 
were placed near GC candidates and we avoided the disk region but otherwise 
sources were distributed at random on the image. We used 
DAOFIND with the same detection parameters to recover the simulated sources.
%The resulting I band completeness function is shown in Fig. **.
Our tests suggest that we are 100\% complete at our cutoff magnitude
of I $\sim$ 23.
%complete to I $\le$ 25. The B band completeness function has a
%similar shape and is complete to B $\le$ 26.5.
This corresponds to M$_I$ $\sim$ --8.5, and hence we have not quite reached the 
expected GC luminosity function turnoff at M$_I$ $\sim$ 
--8.3. In other words, we have 
detected slightly less than half of the GC system in terms of magnitude.

\subsection{Number of GCs and Specific Frequency}

%Our final candidate GC list contains 142 objects. 
As stated above, 
we expect contamination from foreground stars in our final object list to be 
minimal. As the surface
density of objects declines with galactocentric distance, and our spatial coverage is
a fraction of the Milky Way GC system extent, the number of
unresolved background galaxies with colours matching those of GCs is
likely to be very small. From our visual inspection of the images, 
we suspect that the main source of any contamination
is misclassified starforming regions (combined with some internal reddening to
produce colours similar to a GC) located close to the galaxy disk. If
as many as half of the 26 objects with 1 $<$ $|Z|$ $<$ 2 kpc of the
galaxy disk are actually starforming regions, then the resulting uncertainty on our 
final object list of 142 is 9\%. 
%total number of
%candidate GCs reduces to 129. 
This gives a first order estimate of the uncertainty associated 
with misclassified starforming regions. 

It is also possible that {\it bona fide} GCs have been reddened
sufficiently by internal dust so as to shift their colours beyond our
red limit, i.e. (B--I)$_o$ $>$ 2.2. However the small relative number of
red ((B--I)$_o$ $\sim$ 1.9) GCs that we detect suggests that reddening of
intrinsically blue GCs is a minor effect.

In order to calculate the total GC system of NGC 5170, 
we require a factor of two correction in both 
spatial area and luminosity (with an estimated 10\% uncertainty on each). 
Adding the various uncertainties in quadrature gives a total error of 17\%.
Our estimate for the total
GC system for NGC 5170 is 600 $\pm$ 100. 
%The error on this value is
%dominated by our choice of region to exclude above and below the disk,
%which we estimate to be $\pm$ 10\%.  
An additional source of
uncertainty in this estimate is the distance to NGC 5170, which
strongly affects the magnitude incompleteness correction. If the
galaxy is actually at a larger distance (as suggested by the
Tully-Fisher measurements of Willick et al.  1997) then our estimate
of the total number of GCs would be an upper limit.
%The specific frequency is less distance
%sensitive than the total number and f
For a distance of 19.5 Mpc and M$_V$ = --21.6, we calculate a 
specific frequency of S$_N$ = 1.37 $\pm$ 0.23.
This can be compared to the GC system numbers derived by Fischer et
al. (1990) of 815 $\pm$ 320 and S$_N$ = 1.3 for an assumed distance of
31.6 Mpc and M$_V$ = --22.0.

\section{A possible Ultra Compact Dwarf}

Our ACS imaging field includes a bright object which is excluded from
our final GC list. Its magnitude I$_o$ = 19.00 and its size of
0.065$^{''}$ (in both B and I filters) corresponds to
%M$_I$ = --12.79 and r$_h$ = 7.18 pc respectively at a distance of
%22.8 Mpc. For our revised distance of 19.5 Mpc, its luminosity and size
are M$_I$ = --12.45 and r$_h$ = 6.14 pc for a distance of 19.5 Mpc.  It
has a relatively red colour (B--I)$_o$ = 1.89 and it appears quite
round in the ACS images. 
%It has photometric properties similar to Omega Cen. 
It is
located at R.A. = 13:29:53.4 and Dec. = --17:59:09.1 (J2000), SE of the
galaxy disk (see Figures 1 and 5).  A recession velocity is needed to confirm its
association with NGC 5170.

Whether this object should be classified as a massive GC or an Ultra Compact 
Dwarf is a matter of debate. 
The properties of objects classified as UCDs overlap with those that
have been classified as massive GCs (e.g. Forbes et al. 2008). 
Indeed the properties of the object described above  
lie within the range of those found for GCs 
in the rich GC systems studied by Harris (2009a). So although 
the object is clearly distinct from the bulk of the 
NGC 5170 GCs in terms of its 
luminosity, it is still consistent with being drawn 
from a (universal) GC luminosity function.   
Mieske et al. (2006b) have suggested that a GC/UCD 
transition occurs at M$_V$ $\sim$ --11 (M$_I$ $\sim$ --12) or 
a few million solar masses, as this marks the 
change from a near constant size to one that scales with luminosity.  
(Mieske et al. 2006b; Forbes et al. 2008). By this definition, the 
NGC 5170 object would be classified as a UCD.

\begin{figure*}
\begin{center}
%\leavevmode
%\epsfxsize=15cm\epsfbox{size.ps}
\includegraphics[scale=0.25]{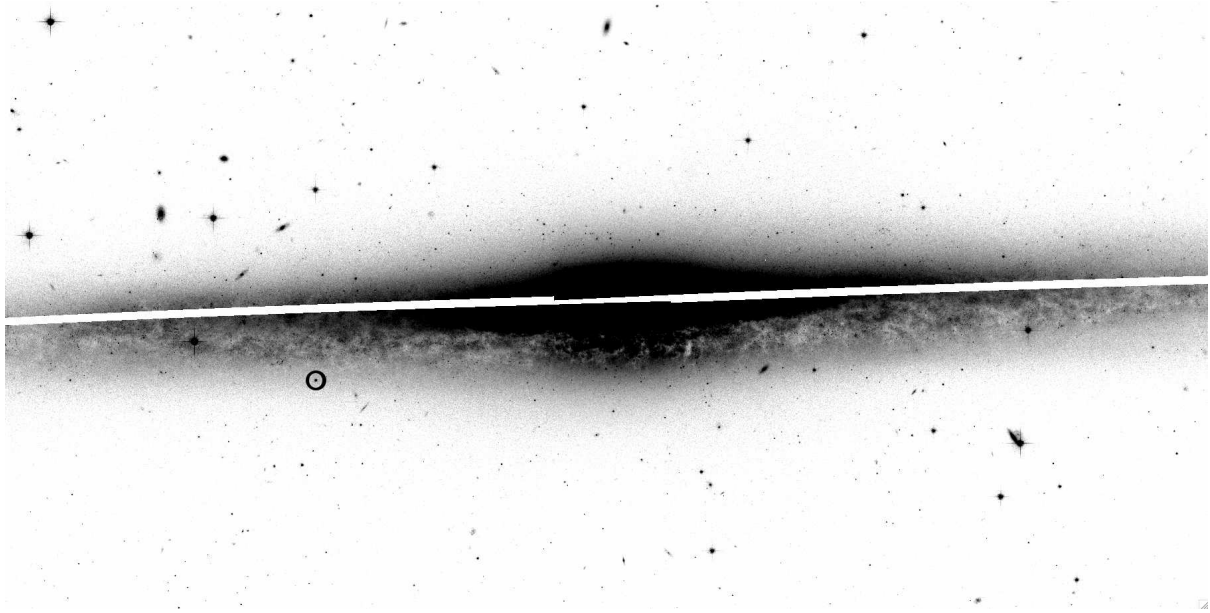}
\caption{HST/ACS I band image of NGC
  5170 covering 370$^{''}$ $\times$ 210$^{''}$. 
North is up and East is left. The inter-chip gap runs across
  the image East-West. The location of a potential Ultra Compact Dwarf is shown by an open circle. 
%Dust (shown white) is clearly seen along the
%  Southern edge of the disk.  
}
\end{center}
\end{figure*}

\begin{figure*}
\begin{center}
%\leavevmode
%\epsfxsize=15cm\epsfbox{size.ps}
\includegraphics[scale=0.5, angle=-90]{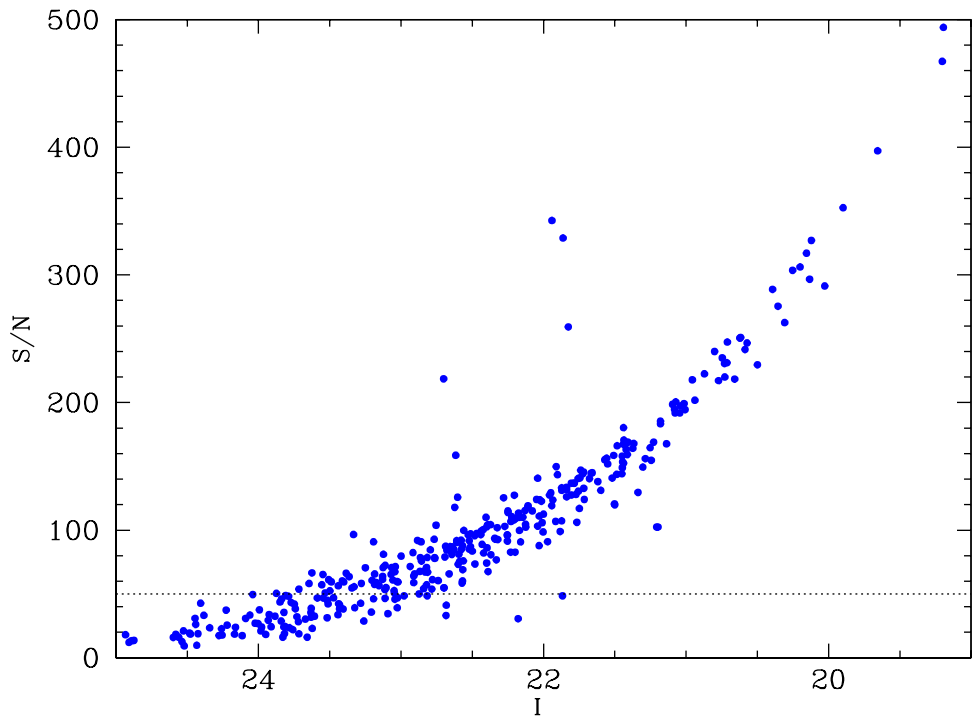}
\caption{I band signal-to-noise ratio as a function of I magnitude. A  
signal-to-noise cut of greater than 50 
(shown by the horizontal line) is adopted for object selection. 
}
\end{center}
\end{figure*}

\begin{figure*}
\begin{center}
%\leavevmode
%\epsfxsize=15cm\epsfbox{size.ps}
\includegraphics[scale=0.5,angle=-90]{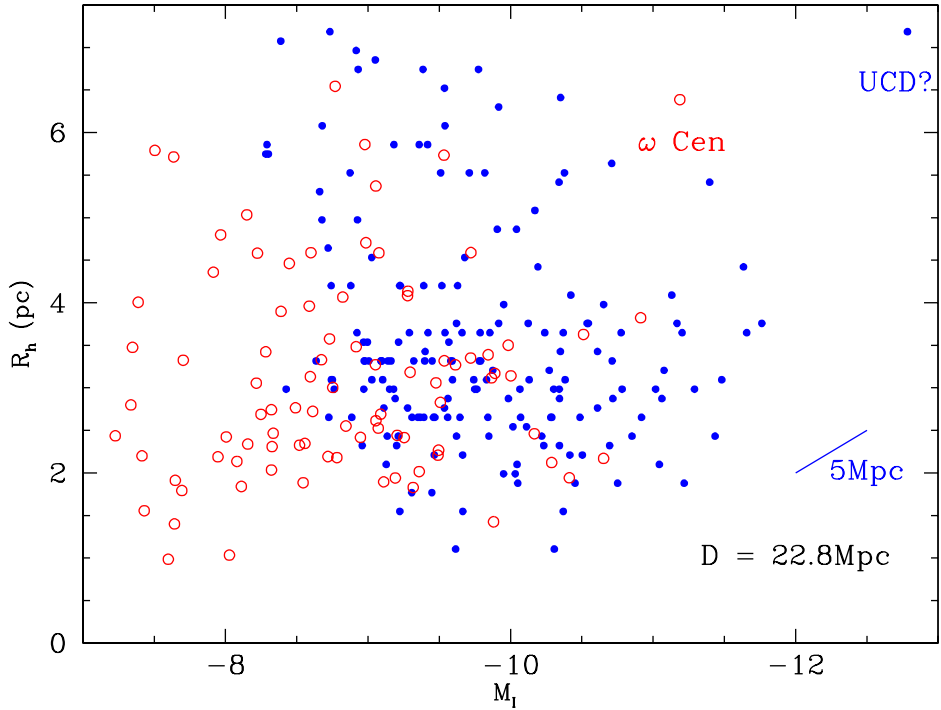}
\caption{Half light radius vs extinction-corrected I band absolute 
magnitude for NGC 5170 candidate GCs (blue filled circles) compared to 
Milky Way GCs (red open circles). A distance of 22.8 Mpc is assumed. 
The length and angle of the solid line indicates the effect of a change of 
5 Mpc in distance to NGC 5170. The location of Omega Cen and the 
possible Ultra Compact Dwarf around NGC 5170 are labelled. The locus of 
NGC 5170 objects is offset from that of the Milky Way GCs. 
}
\end{center}
\end{figure*}

\begin{figure*}
\begin{center}
%\leavevmode
%\epsfxsize=15cm\epsfbox{sizebest.ps}
\includegraphics[scale=0.5,angle=-90]{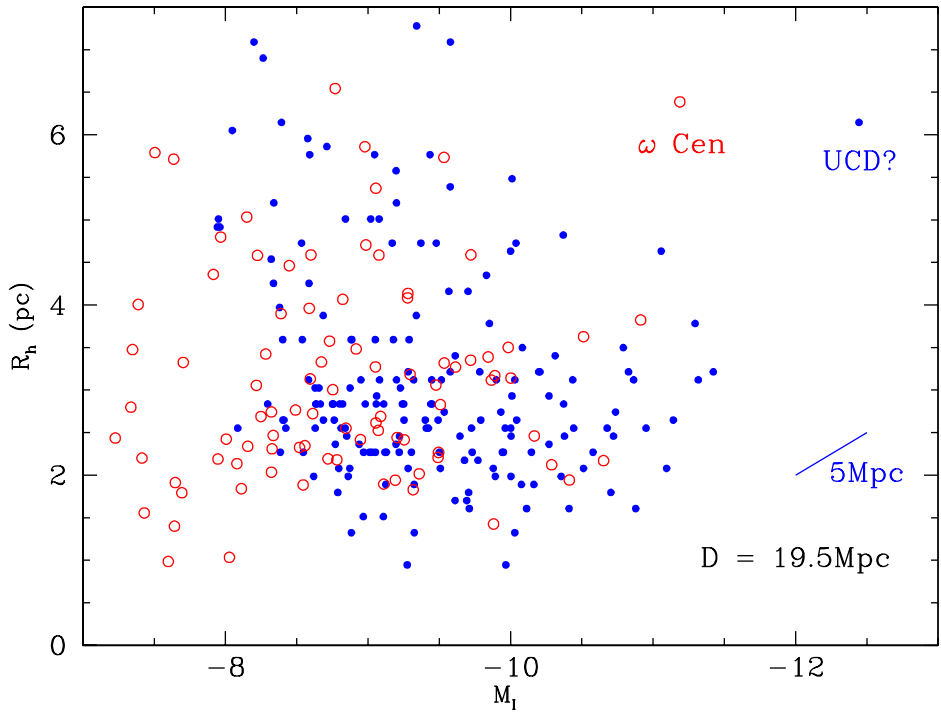}
\caption{Half-light radius vs extinction-corrected I band absolute 
magnitude for NGC 5170 candidate GCs compared to 
Milky Way GCs. A distance of 19.5 Mpc is assumed. Other labels as per 
Fig. 3. 
%The length and angle of solid line indicates the effect of a change of 
%5 Mpc in distance to NGC 5170. 
The locus of NGC 5170 objects is 
similar to the Milky Way size-magnitude distribution for a distance of 
19.5 Mpc. 
}
\end{center}
\end{figure*}

\begin{figure*}
\begin{center}
\includegraphics[scale=0.5,angle=-90]{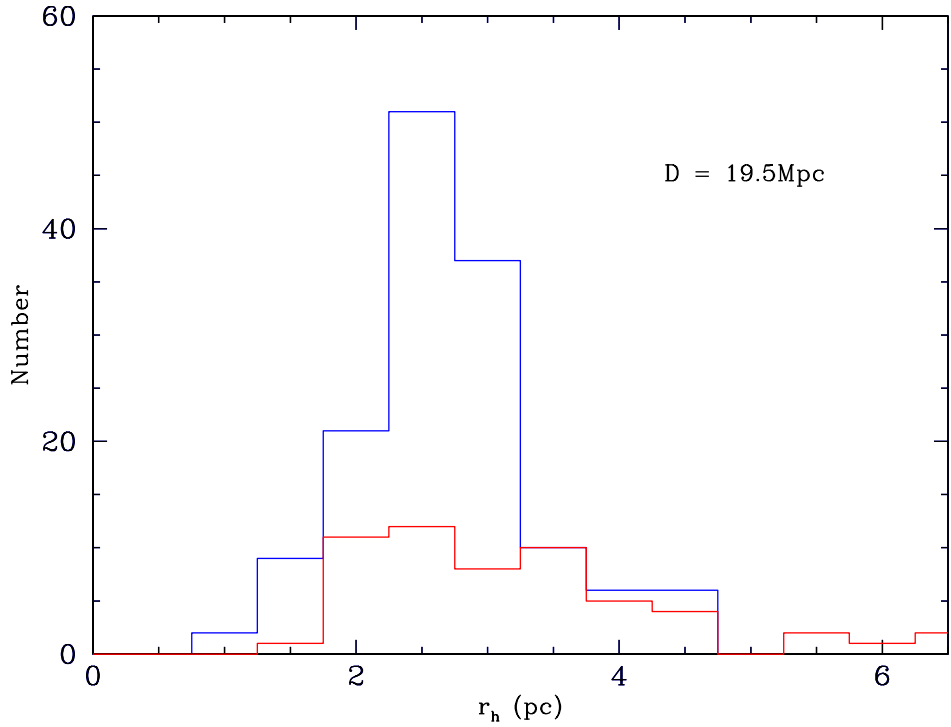}
\caption{Histogram of half-light radius. The upper blue histogram shows the 
NGC 5170 candidate GCs, while the lower red histogram shows the Milky Way GC 
system restricted to M$_I$ $<$ --8.5.   
A distance of 19.5 Mpc is assumed. The two distributions have the same 
median half light radius. 
}
\end{center}
\end{figure*}

\begin{figure*}
\begin{center}
\includegraphics[scale=0.5,angle=-90]{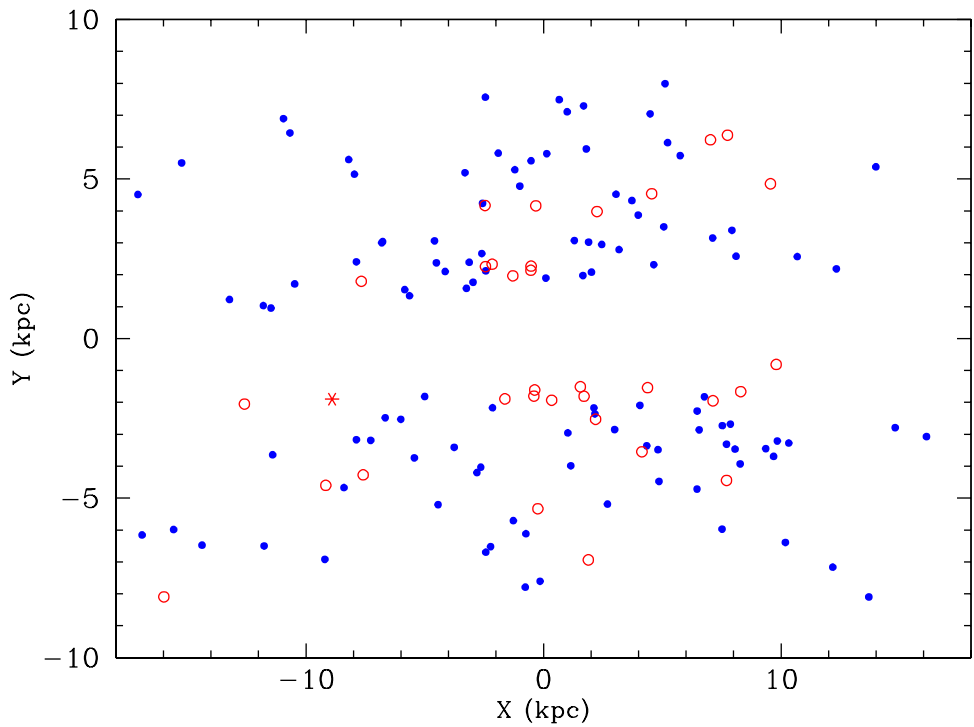}
\caption{Spatial location of GC candidates around NGC 5170. 
The bulge and disk regions have been excluded from 
the selection process. Blue GCs (B--I $<$ 1.75) are shown by blue filled circles, and red GCs by red open circles. 
The location of the possible Ultra Compact Dwarf 
is shown with a red star symbol. A distance of 19.5 Mpc is assumed. 
}
\end{center}
\end{figure*}

\begin{figure*}
\begin{center}
%\leavevmode
%\epsfxsize=15cm\epsfbox{cmd.ps}
\includegraphics[scale=0.5,angle=-90]{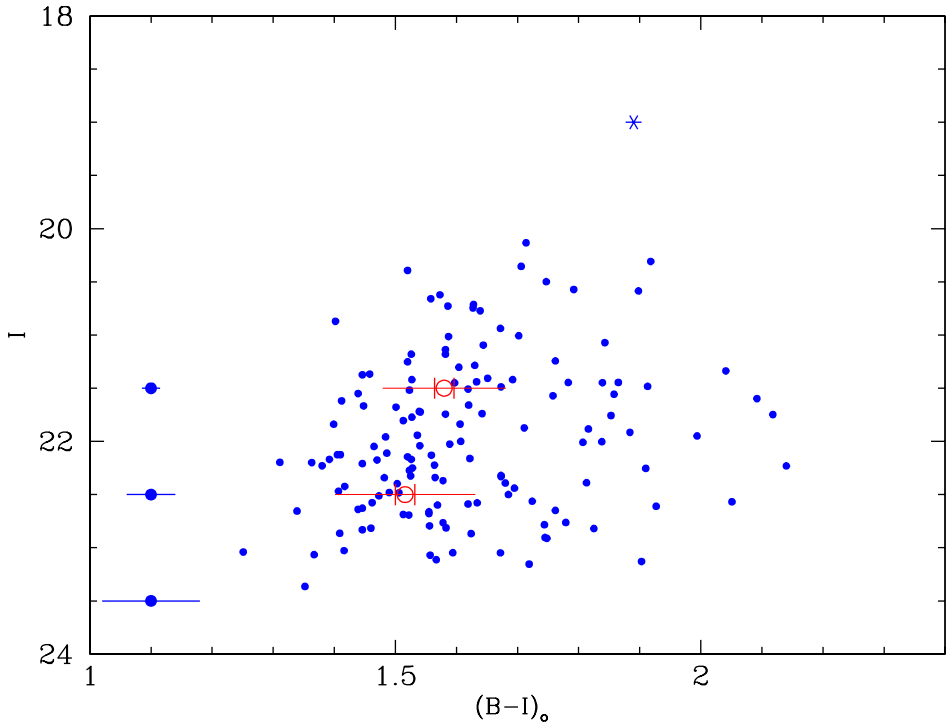}
\caption{Observed colour-magnitude diagram for NGC 5170 candidate GCs. 
The mean value (open circle), error on the mean (vertical tick) 
and rms dispersion (horizontal line) for the blue subpopulation from the 
NMIX statistical test are overlaid. 
A colour difference is seen, at the $\sim$3$\sigma$ level, 
between the blue GCs brighter/fainter than I = 22, i.e. a blue tilt.
%A blue tilt can be seen in the blue (B--I $\sim$ 1.6) 
%subpopulation of GCs. 
The red (B--I $\sim$ 1.9) 
subpopulation is sparse as we have partially selected against bulge/disk GCs.
%Size selection of r$_h$ $<$ 5.5 pc applied, and subsample of objects 
%with r$_h$ $<$ 3.9 pc (blue). 
}
\end{center}
\end{figure*}

\begin{figure*}
\begin{center}
%\leavevmode
%\epsfxsize=15cm\epsfbox{cmd.ps}
\includegraphics[scale=0.5,angle=-90]{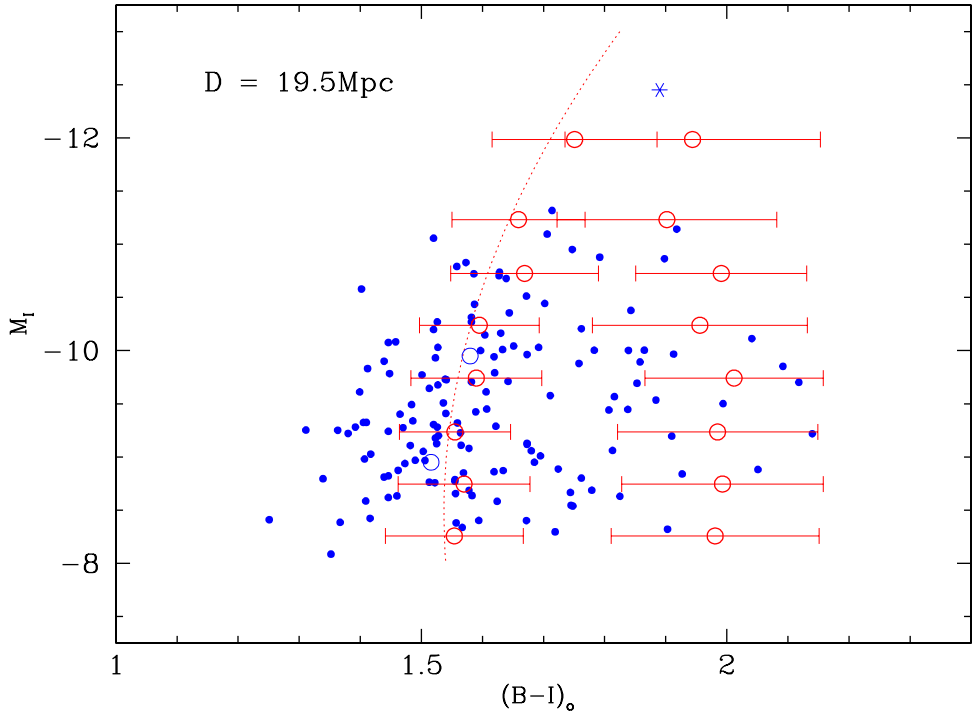}
\caption{Colour-absolute magnitude diagram for NGC 5170 candidate GCs 
(assuming a distance of 19.5 Mpc). Also included is the potential UCD, given by a star 
symbol. 
Red open circles and error bars represent the mean values and rms dispersions for the 
combined sample of massive ellipticals from Harris (2009a). A 
2nd order polynomial fit to the massive elliptical data is given by a dashed line.
Blue open circles without error bars are the mean values for the NGC 5170 blue GC 
subpopulation.  
In the Milky Way, 
Omega Cen has M$_I$ = --11.2 and the (universal) GC luminosity function 
turnover is at M$_I$ $\sim$  --8.3. 
The blue subpopulation in NGC 5170 is consistent 
with the blue tilt seen in  
the massive ellipticals but covers a much smaller magnitude 
range with fewer GCs. 
}
\end{center}
\end{figure*}

\begin{figure*}
\begin{center}
%\leavevmode
%\epsfxsize=15cm\epsfbox{cmd.ps}
\includegraphics[scale=0.5,angle=-90]{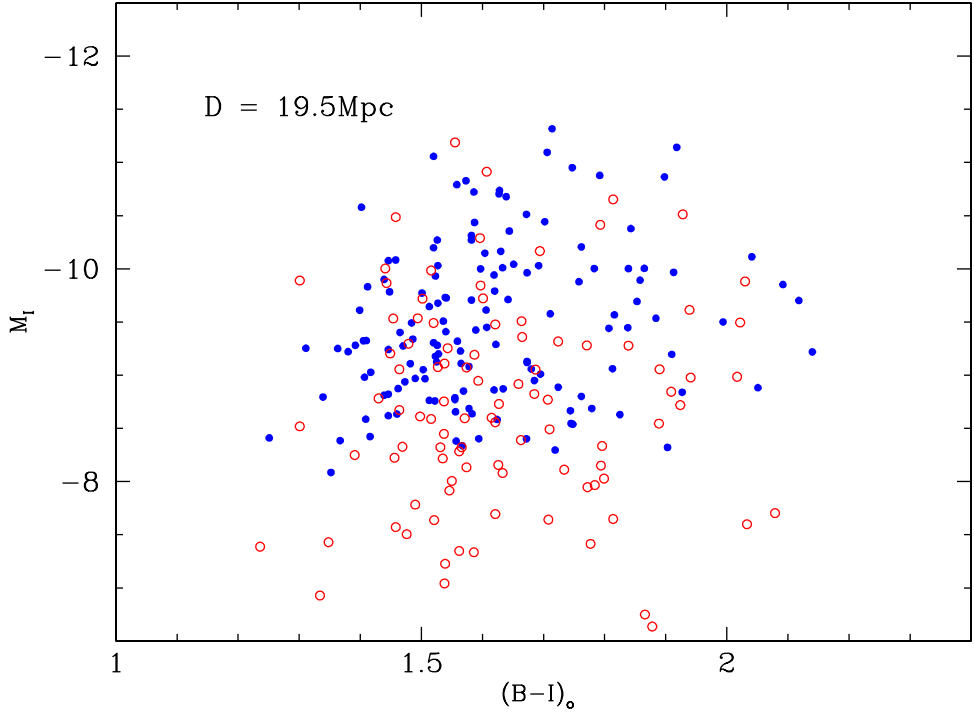}
\caption{Colour-absolute magnitude diagram showing 
NGC 5170 candidate GCs (blue filled circles) and Milky Way GCs (red 
open circles). The colour-magnitude 
distributions are similar for the two GC systems, although 
NGC 5170 clearly includes more objects brighter than our 
cutoff magnitude of M$_I$ $\sim$ --8.5. 
}
\end{center}
\end{figure*}

\begin{figure*}
\begin{center}
%\leavevmode
%\epsfxsize=15cm\epsfbox{cmd.ps}
\includegraphics[scale=0.5,angle=-90]{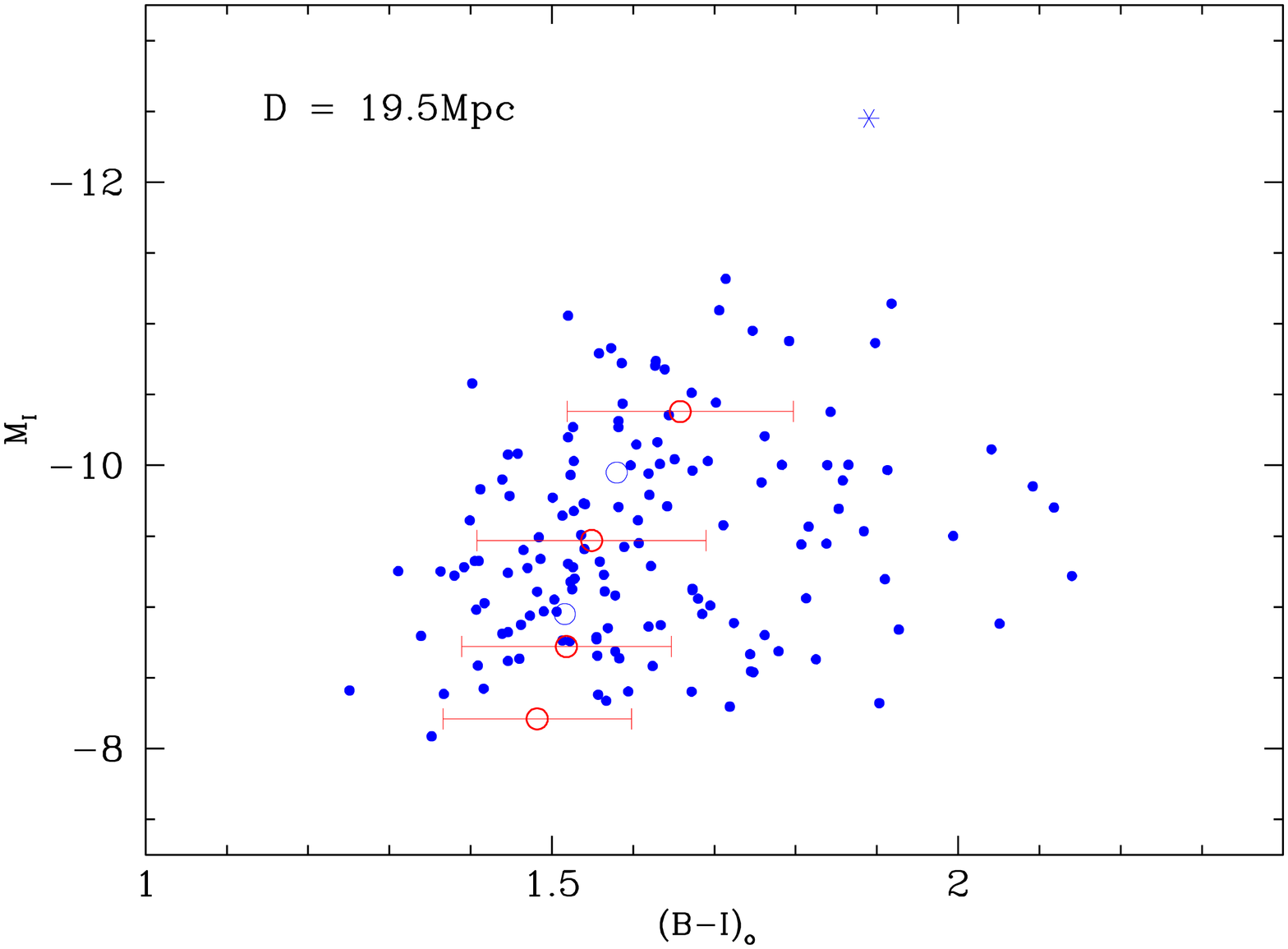}
\caption{Colour-absolute magnitude diagram showing 
NGC 5170 candidate GCs 
(small blue circles) with mean values and rms dispersion for the  
simulated data (large red circles with error bars).  
Blue open circles without error bars are the mean values for the NGC 5170 blue GC 
subpopulation.  
Also included is the potential UCD, given by a star 
symbol. 
The simulated data are generated from the self enrichment model, 
which includes dynamical mass loss,  
as described in Harris et al. (2009a). 
%The mean 
%values and rms dispersions for the simulated data 
%are consistent with the 
%blue subpopulation GCs in NGC 5170.
%with the blue tilt seen in the simulated data at bright magnitudes.   
}
\end{center}
\end{figure*}

\section{Discussion and Conclusions}

Using two pointings of the HST/ACS camera, in the B and I filters, we
have explored the GC system surrounding the edge-on spiral galaxy NGC
5170. We employed the aperture-correction method of Harris (2009a) for
partially resolved objects to obtain accurate total magnitudes.  The
initial selection of candidate GCs is based on (B--I)$_o$ colour and
signal-to-noise ratio (which closely tracks magnitude).  We used the
object size, as measured by ISHAPE, to further refine the selection
process and to provide an estimate of the distance to NGC 5170
(under the assumption of a universal average size for GCs). This 
distance is 19.5 Mpc and is within $\sim$2$\sigma$ of the Hubble flow 
distance.

The colour-magnitude diagram reveals a well-populated blue subpopulation and a sparse 
red subpopulation, in a ratio of roughly 3 to 1. The GC subpopulation 
colours are similar to those seen in the Milky Way and M31 GC systems.
We also find a  
blue tilt that is similar, 
within our restricted magnitude range, 
to that for massive ellipticals found by Harris (2009a) and to the 
self enrichment model described in Harris et al. (2009a). 
These findings suggest that normal late-type spiral galaxies, along with
galaxies of earlier types, host GC systems with a blue tilt.  The blue
tilts are most likely evidence for a mass-metallicity relation 
%of the form Z $\sim$ L$^{\alpha}$ (where 
%$\alpha$ $\sim$ 0.3, Z is the total metallicity and L is the 
%luminosity) 
within the metal-poor GC subpopulation. The reason for the lack of an
observed blue tilt in the Milky Way GC system appears to be simply due
to the small number of relatively massive GCs 
(Bailin \& Harris 2009; Harris 2009a). 
We would expect that new, high precision
photometry of a large sample of the M31 GC system should reveal a blue tilt.

We note
that the reality of blue tilts in general has been disputed by Kundu
(2008), and in the case of M87 in particular by Waters et
al. (2009). Subsequent work by Harris (2009a,b), 
Peng et al. (2009) and Madrid et
al. (2009) has addressed the issues raised by these workers
and has strongly confirmed the astrophysical reality of blue tilts. 
%However, it is a subtle 
%effect which requires accurate photometry.

Although of similar Hubble type and total magnitude  
to the Milky Way (Sbc, M$_V$ $\sim$ --21.3), we find that the GC
system of NGC 5170 is much richer with a total GC population of 600
$\pm$ 100 GCs. In this sense NGC 5170 bears more resemblance to the Sb
galaxy M31 with $\sim$450 GCs (Barmby et al. 2000), 
or the Sa Sombrero galaxy with 1900 GCs (Rhode \& Zepf 2004). 

A commonly-used measure to directly 
compare GC systems is the specific frequency S$_N$, which normalises GC number by M$_V$. 
Assuming M$_V$ = --21.6 for NGC 5170, we estimate S$_N$ = 1.37 $\pm$ 0.23.  This is higher 
than the Local Group spirals (i.e. for the Milky Way S$_N$ = 0.6 and M31 S$_N$ = 0.9) and 
the four Sb-Sc spirals studied by Rhode et al. (2007)
%, they found total GC systems of 80
%to 290 and a corresponding specific frequency 
who found a range of 0.5 $<$ S$_N$ $<$ 0.9, but less than S$_N$ = 2.1 for the Sombrero 
galaxy (Rhode \& Zepf 2004). 
%respectively (Ashman \& Zepf 1998). For the Sc galaxy NGC
%3370, Cantiello et al. (2009) estimated a lower limit of 138 GCs (and
%a specific frequency of S$_N$ $>$ 0.8). In the ground-based study of
An alternative measure for comparing GC systems 
is the stellar mass normalised GC frequency,   
T$_N$ (Rhode \& Zepf 2004; Spitler et al. 2008). 
Using the 2MASS K band magnitude of 7.63 and a Salpeter IMF we derive a stellar mass of 7.6  
$\times$ 10$^{10}$ M$_{\odot}$. This gives T$_N$ $\sim$ 8. However this likely  
an upper limit given that some fraction of the K band flux will be obscured due to 
the highly edge-on ($i$ $\sim$ 86$^o$) nature of NGC 5170. 
Spitler et al. showed that T$_N$ varies with the stellar mass of the host, and that 
the Sombrero galaxy had the highest T$_N$ value ($\sim$ 9) of the 8 spiral galaxies listed. 
Thus NGC 5170 appears to have a high GC frequency, placing it between 
the Local Group spirals and the Sombrero galaxy.

Using the scaling relation of Spitler \& Forbes (2009) between 
total GC system mass with halo mass, we estimate 
a halo mass for NGC 5170 to be 3.4 $\times$  10$^{12}$ M$_{\odot}$. This is a factor of 
2-3$\times$ that of the Milky Way halo mass.

We also report the discovery of a possible Ultra Compact Dwarf 
associated with NGC 5170.  If confirmed by its radial velocity, it
will be another rare example of a UCD located near a spiral galaxy
outside that of a galaxy cluster (see Hau et al. 2009).

\section*{Acknowledgments}

We would like to thank 
K. Forde for his work performing the initial data reduction. 
DF and LS thanks the ARC for financial support. 
We thank C. Foster for useful comments on this work. 
This material is based upon work supported by NSF grant AST-0507729. We 
acknowledge the usage of HyperLeda and NED databases. Finally, we thank the referee for 
several useful suggestions that have improved the paper.

\end{document}